\documentclass[journal]{IEEEtran}
\usepackage{cite}
\usepackage{graphicx}
\usepackage{amsmath, amssymb}
\usepackage{subcaption}
\usepackage{url}

\begin{document}
\title{Unsupervised adulterated red-chili pepper content transformation for hyperspectral classification}

\author{Muhammad~Hussain~Khan,
        Zainab~Saleem, 
        Muhammad~Ahmad,
        Ahmed~Sohaib,
        and~Hamail~Ayaz
\thanks{M. H. Khan, Z. Saleem, M. Ahmad, A. Sohaib, and H. Ayaz are with the Department of Computer Engineering, Khwaja Freed University of Engineering and Technology (KFUEIT), Rahim Yar Khan, 64200, Pakistan. e-mail: mahmad00@gmail.com}}
\maketitle
\begin{abstract}
Preserving red-chili quality is of utmost importance in which the authorities demand the quality techniques to detect, classify and prevent it from the impurities. For example, salt, wheat flour, wheat bran, and rice bran contamination in grounded red chili, which typically a food, are a serious threat to people who are allergic to such items. This work presents the feasibility of utilizing visible and near-infrared (VNIR) hyperspectral imaging (HSI) to detect and classify the aforementioned adulterants in red chili. However, adulterated red chili data annotation is a big challenge for classification because the acquisition of labeled data for real-time supervised learning is expensive in terms of cost and time. Therefore, this study, for the very first time proposes a novel approach to annotate the red chili samples using a clustering mechanism at $500~nm$ wavelength spectral response due to its dark appearance at a specified wavelength. Later the spectral samples are classified into pure or adulterated using one class SVM. The classification performance achieves $99\%$ in case of pure adulterants or red chili whereas $85\%$ for adulterated samples. We further investigate that the single classification model is enough to detect any foreign substance in red chili pepper rather than cascading multiple PLS regression models.
\end{abstract}
\IEEEpeerreviewmaketitle
\section{Introduction}

\IEEEPARstart{H}{yperspectral} Imaging (HSI) is leading-edge technology which considers a wide electromagnetic spectrum range of light instead of just primary colors made from the visible range such as red, green, and blue to characterize a pixel \cite{Mahmad2019A}. The light striking a pixel, indeed, is divided into many different spectral bands to provide more detailed information on what is imaged. HSI provides not only the spatial information (\textit{Shape of object}) of the object but also its spectral information (\textit{Type of material and chemical distribution of elements with which it is composed of}). HSI exists in a 3D cube form (\textit{Hyper-Cube}) containing all the spatial images in the form of stack with respect to its wavelength spectrum having coordinates $x,~y,~\lambda$, where $x$ and $y$ are spatial coordinates and $\lambda$ is the spectral wavelength coordinate and thus each pixel of a Hyper-Cube can be interpreted as an individual spectrum.

HSI systems are usually classified in two regions: NIR (near infra-red) \& MIR (mid infra-red). NIR region ($780-2500~nm$) detect patterns which reveals chemical and physical combination of materials while MIR region ($2500-25000~nm$) describes rotational and vibrational motion of molecules which are highly sensitive to composition of materials \cite{infrared}. However penetration power of NIR radiations (up to several millimeters) is superior than MIR (in microns) which effects reliability and precision of multivariate analysis \cite{penetration}.

HSI has been adopted in a wide range of real-world applications including food sciences, biomedical imaging, geosciences, forensic and surveillance to mention a few \cite{Mahmad2019}. One of the main challenges in the HSI domain is the characteristics of the data, which typically yields hundreds of contiguous and narrow spectral bands with very high spatial resolution throughout the electromagnetic spectrum \cite{Mahmad2017}. Therefore, HSI classification (HSIC) is complex and can be dominated by a multitude of classes and nested regions, than the traditional monochrome or RGB images \cite{MAhmad2019B}.

HSI has been used in remote sensing for years but recently it has become an important and robust tool to collect information about how the light reacts with different materials and thus has been widely used in food and agriculture quality control. For example, characterizing the quality of food \cite{moisture} or detection of adulterants in various food items \cite{al2018detection}. Quality is of utmost importance in food industry, where the industry not only demands traditional techniques like preservation of food by freezing and cooling, detection of deceased food items and adulterants, it also demand modern and convenient methods to ensure the quality of food.

Adulteration can be defined as addition of constituents in food items which is forbidden by law, customs, norms and practices. Deception is one of the most frequently used practice internationally. Adulteration in food items was first reported by Thephrastus $(370-285~BC)$ and effort to address this problem traces back to Roman civil law. Although, the reasons behind adulteration of foods are mostly economical or financial, it can lead to serious public health concerns. For examples, in $1994$, powdered paprika was found to be adulterated with lead oxide which caused death of several people in Hungary \cite{asta}.

\begin{table*}[hbt]
  \centering
  \caption{Example of spectroscopic method for detection of adulterants}
    \begin{tabular}{c|c|c|c}
    \textbf{Product} & \textbf{Adulterant} & \textbf{Method} & \textbf{Reference} \\ \hline
    Saffron & Saffron, Mrigold, and Turmeric & FT-NIR, Raman, and LIBS & \cite{saffron} \\ \hline
    Onion Powder & Cornstarch & FT-NIR & \cite{Onion} \\ \hline
    Chili Powder & Sudan Dyes & Raman and NIR & \cite{SudanDyes} \\ \hline
    Tumeric & metanil yellow and Chalk Powder  & FT-NIR and Teragertz Spectroscopy & \cite{Tumeric} \\ \hline
    Black Pepper & Millet and Buckwheat & NIR & \cite{Blackpepper} \\ \hline
    Paprika & Tomato Skin and Brick Dust & FT-NIR & \cite{galaxy}  \\ \hline
    \end{tabular}
  \label{Tab.1}
\end{table*}

Red pepper is a fruit of capsicum anum and widely used in many cuisines as spice \cite{korean}. Usually red pepper is used in powdered or crushed form, made by drying and grinding or crushing the ripped fruit. However, due to high prices of red chili, many inexpensive material are added by vendors to make more profits. Now a days, food a are charging on the basis of purported reports without any mechanism of testing the red chili pureness on run time. The experts mostly rely on observations or the samples are sent to distant laboratories to conduct testing for more accurate results. Traditional methods that has been widely used for the detection of adulteration in food items are Liquid chromatography \cite{LC}, isotope ratio mass spectrometry \cite{Recent}, gas chromatography \cite{DataBase}, and hyphenated mass spectroscopy \cite{DataBase}. All these method requires skilled analyst and careful analysis which can take days to conclude the results. This obstruction commonly creates a conflict between food authorities and vendors.

Recently, in pursuit of rapid detection of adulterants in food items, fingerprinting of human food using vibrational spectroscopy techniques  \cite{Recent} such as Fourier Transform near-infrared (FT-NIR) \cite{galaxy}, near-infrared (NIR) \cite{NIR}, Raman \cite{Raman}, and visible near-infrared (VNIR) spectroscopy \cite{korean} is gaining interest of many researchers and industries. NIR spectra arises due to weak and broad molecular bonds and overtones, mostly associated with methine, hydroxy, and amine functional groups \cite{FT}. By using interferogram and Fourier transform techniques in acquisition of spectra, FT-NIR spectroscopes have enhance performance in wave number precision and reproducibility as compared to dispersion NIR \cite{FT}. However, the choice of instrument used is dependent on the application while considering many vital aspects like spatial and spectral resolution, accuracy and wavelength range. To detect adulterant in food items, a number of methods based on spectral imaging and chemomatric analysis has been proposed in last few decades \cite{Recent}. Table \ref{Tab.1} enlist few state-of-the-art works with the focus on adulteration identification in food items. 

Capsaicionoids are key components of red chili, cause sensations of burning in tissues when comes in contact. Red Chili quality is graded on the basis of capsaicionoids content \cite{hot}. Jongguk Lim, et al. develop a system which determines capsaicionoids content in Korean red pepper powder. They used Visible and Near Infra Red (VNIR) spectrometer ($450-950~nm$) along with first order derivative pre-treatment method and Partial Least-Squares Regression (PLSR) model to predict amount of capsaicionoids content in red paprika powder \cite{korean}. In another work, they also detect the moisture content in grounded red chili of various particle sizes by using NIR spectroscopy combined with PLSR model and concluded that the performance can be enhanced by limiting the range of particle size \cite{korean1}.

Aflatoxin are poisonous carcinogen produced by aspergillus. Aspergilus is common in warm, humid environment with native habitat in soil and it permeate in organic matter whenever conditions are conducive for its growth \cite{Aflatoxin}. As red chili need to be dried before storing, crushing and grinding, the poor sanitary measures may results in contamination with soil born diseases like aflatoxin. There are various instance globally which reported presence of $aflatoxin-B1$ in red chili \cite{mycotoxin, mycotoxin1}. H. Kalkan, et al. used multispectral imaging system with effective wavelength of $400-510~nm$ to detect aflatoxin contamination in red pepper and hazelnut. They used a modified three dimensional version of linear discriminant analysis (LDA) algorithm for generation and selection of features. The developed algorithm were able to classify contaminated pepper flake with uncontaminated with accuracy of 79.17\% \cite{kalkan2011detection}. Smita Tripathi and H.N. Mishra proposed a FT-NIR based method along with PLSR to detect the presence of $aflatoxin-B_1$ ranging from $15$ to $500~\mu~g/kg$. They compared their result with techniques like high performance liquid chromatography and thin layer chromatography and postulate that developed FT-NIR based system's efficiency is comparable with traditional arduous chemical methods \cite{tripathi2009rapid}. 

As earlier discussed, red chili is preserved by drying. Traditionally, red chili is dried in sun for $7$ to $20$ days which increase risk of contamination with fungal disease, foreign particles, etc \cite{characterization}. Another method used for drying red chili is hot air ($45-70^\circ C$) which is more convenient and hygienic \cite{tunde}. However, In Min Hwang, et al., found that the sun dried chili price is $30\%$ higher than hot air dried. Hence, they characterize the red pepper by using  High Performance Liquid Chromatography (HPLC) and NIR spectroscopy with LDA and found out that sun dried chili have slightly higher values of American Spice Trade Association (ASTA) color values, free sugar, lactic acid and capsaicin \cite{characterization}. Xi-YU Wu, et al. used NIR spectroscopy coupled with PLSR to detect commonly used adulterants (wheat bran, rice bran, rosin powder, and corn flour) in sichuan pepper powder. They mixed different quantities of adulterants in sichuan pepper powder and predicted different composition with prediction coefficients $0.971, 0.948, 0.969, 0.967, 0.994$ for Sichuan pepper powder, rice bran, wheat bran, corn flour and rosin powder respectively \cite{similiar}.

Most of the efforts in past utilizes HSI with PLSR for detection of adulterants in powdered materials \cite{PLS1}. However, PLSR is facing a lot of criticism in research community. Antonakis et al. suggest and encourage researcher to abandon PLS \cite{PLS2} while Rönkkö et al. recommended to avoid use of PLS as it is extremely tough to justify results \cite{PLS3}. In another paper Rönkkö et al., proposed a ban on the use of PLS \cite{PLS4}. Beside the criticism, PLSR worked on finding the strongest relationship between provided information and PLS factors which is highly dependent on the type of adulterant added. Hence, a separate model is needed for each and every adulterant. It has also been noted that spectra for different varieties of same substance results in different variance which is explained by different number of PLS components, thus separate models are required for each variety. Previously, researcher considered and limited their models to commonly used adulterants and specific type of material, while in real-life scenario the types of adulterants cannot be limited and it is difficult to identify the between varieties specifically in powdered form.

Hence, in this study we are proposing a novel technique based on HSI system with wavelength of range $395-1000~nm$ combined with different basic hyper-Cube processing techniques and one class Support Vector Machine (SVM) to detect unhygienic adulterants in grounded red chili. The proposed methodology works in three steps; first, the data has been labeled by exploiting the International  Commission  on Illumination (CIE) standard by annotating the class labels to chili and the adulterant materials. The characteristic red color of chilli powder appears to be the darkest in the wavelength range of $450-550~nm$, therefore in spectral response of mixed sample at $500~nm$, the bright pixels are of adulterant. Second, to remove the curse of dimensionality effect, PCA has been before building the model. finally, a single class SVM model has been trained on pure red chili and tested with adulterated samples. Most of the domestic users and food authorities are only interested in knowing whether the spices are pure or adulterated, with little or no interest in the type of adulterants added. Hence this model will be able to suffice their requirements to some extent. 

The remainder of the paper is structured as follows: Experimental datasets section \ref{sec2} details the information of samples and their preparation, acquired for this study. Methodology section \ref{sec3} presents the theoretical aspects of data acquisition, spectral pre-processing, data annotation (one of the novel approach of this study), reducing the hyper-cube dimension and one-class SVM. Results and Discussion section \ref{sec4} presents the experimental evaluation of our proposed techniques and discusses opportunities and obstacles of model. Finally, conclusion section\ref{sec5} summarizes the major contributions of this study and potential future research directions that can derived from this work.

\section{Experimental Datasets}
\label{sec2}

\begin{figure}[hbt]
\begin{subfigure}{.5\textwidth}
  \centering
  \includegraphics[width=.9\linewidth]{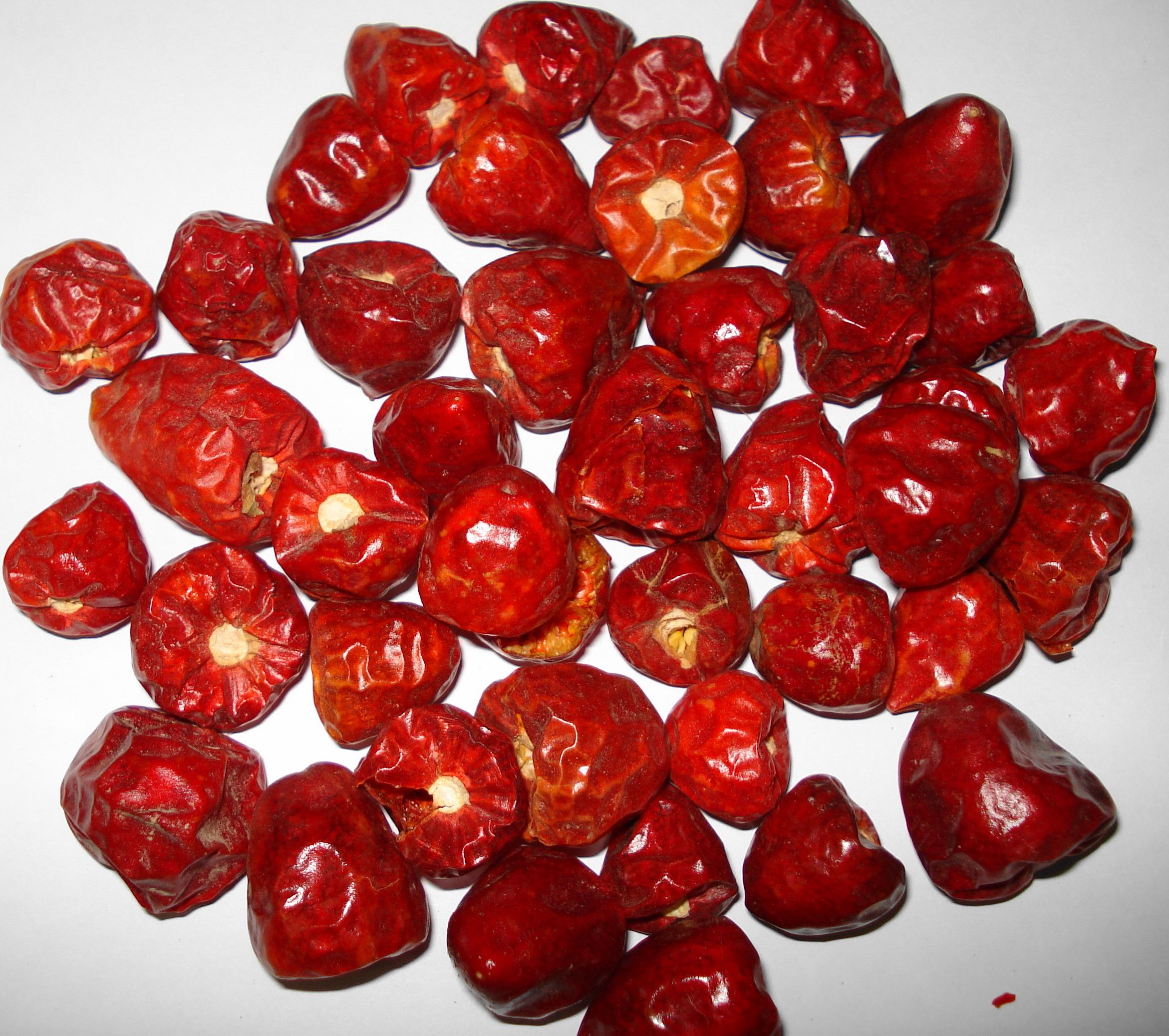}  
  \caption{Kunri; Sindh, Pakistan}
  \label{Fig.1A}
\end{subfigure} \\ \linebreak 

\begin{subfigure}{.5\textwidth}
  \centering
  \includegraphics[width=.9\linewidth]{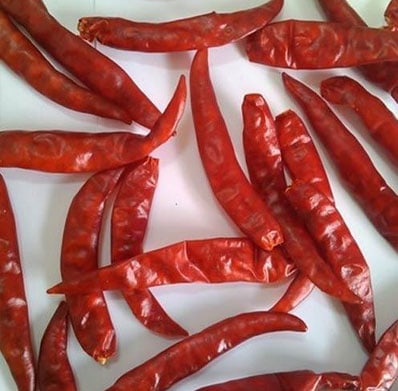}  
  \caption{Hybrid; Rajhastan; India}
  \label{Fig.2A}
\end{subfigure}
\caption{Whole Chili samples analyzed in the study.}
\label{fig:RC}
\end{figure}

In this study, experiment has been carried out on two types of chili samples grown in different origins (i.e, Kunri; Sindh, Pakistan, and Hybrid, Rajhastan, India) and are collected from open market. The red chili samples used in this study are shown in Fig. \ref{fig:RC}. To use red chili in cuisines, it has to be crushed multiple times to make it a fine powder. To make the model consistent, the similar crushing procedure has been replicated for this study. However, during this milling process the chili may get overheat due to which it can loose its natural color. To overcome the aforementioned issues, the samples were cooled at room temperature multiple times during the milling process to preserve the original texture of chili. Three commonly used adulterants wheat bran, rice bran, and saw dust were acquired from local market and cleaned from foreign matter. 

Samples were prepared by adding adulterants individually to both types of chili in range from $0$ to $30\%$ with $2\%$ increment by weight, respectively. Both red chili and adulterants were weighed separately with an electronic balance and mixed by using National mixer grinder to obtain homogenized samples. A total of $54$ samples with $6$ pure chili ($3$ of each origin), $6$ pure adulterants ($2$ of each adulterant), and $42$ adulterated samples ($14$ of each adulterant) were prepared. The samples were seal, packed, labeled and kept at room temperature in order to protect from humid environment. 

\section{Methodology}
\label{sec3}
In this section, first we describe the mode for data acquisition for this study and preprocessing techniques, reflectance calculation, smoothing, scatter removing and dimensionality reduction, needed for spectral evaluation. Further we have discussed our two novel approaches of this research i.e; data annotation and usage of one-class SVM classifier for identifying the adulterant in red chili.

\subsection{Data Acquisition}
\begin{figure}[hbt]
\begin{subfigure}{0.5\textwidth}
  \centering
  \includegraphics[width=0.8\linewidth]{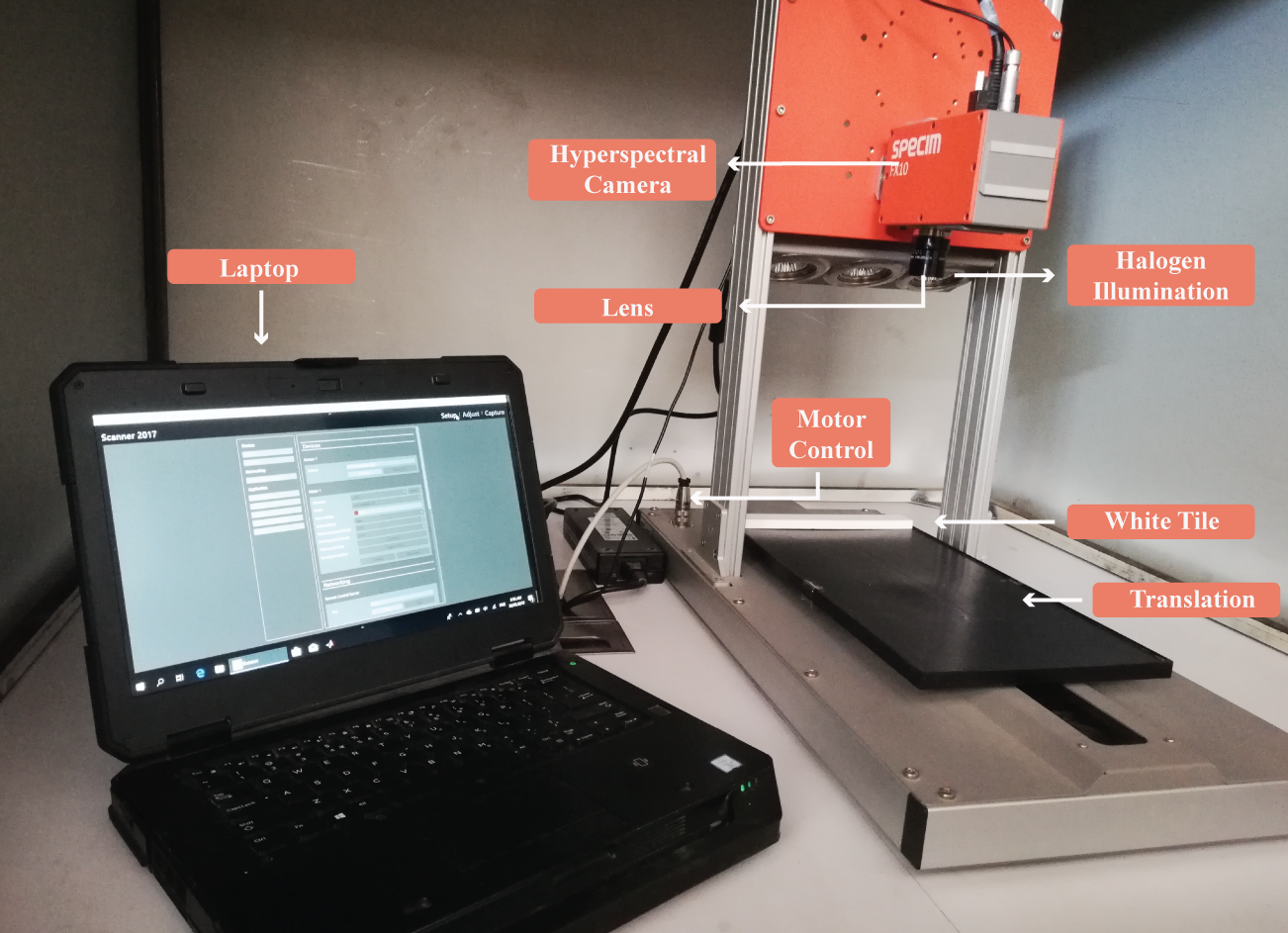}  
  \caption{VNIR Hyperspectral Imaging System}
  \label{HSI}
\end{subfigure} \\ \linebreak

\begin{subfigure}{0.5\textwidth}
  \centering
  \includegraphics[width=0.7\linewidth]{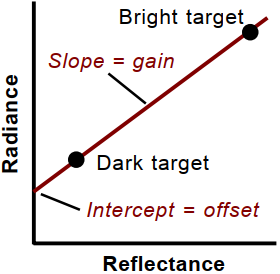}  
  \caption{Empirical Line method for reflectance conversion \cite{van1994calibration}}
  \label{ELM}
\end{subfigure}
\caption{Data acquisition apparatus and reflectance calculation method.}
\label{fig:HSI}
\end{figure}

The HSI system used in this study is shown in Fig. \ref{fig:HSI} which consists of a hyperspectral camera FX-10 (Specim, Spectral Imaging Ltd, Finland) coupled with lens (Scheiner Cinegon $1.4 / 8mm$). The camera is mounted on a lab scanner system which consists of three halogen lamps and a moving platform ($21~cm \times 40~cm$) operated by stepper motor. A computer (Dell-P46g) is connected to the camera through GigE-Vision and scanner via serial communication port. The camera has $224 ~(Spectral ~channels) \times (1024 \times 512) ~(Spatial)$ resolutions. The complete system is sealed in a dark box to avoid ambient noise. All the samples (pure \& adulterated) are placed in petri dishes and leveled by surface leveler to obtain uniform surface and imaged separately. Each sample was scanned at constant speed of $20~m/s$ with exposure time $9~ms$. The acquired hyper-Cube consist of $224$ spectral images representing electromagnetic spectrum (radiance) of scanned materials at different wavelengths spanned from $400-1000~nm$.

The encoded radiance in acquired Hyper-Cube was converted to reflectance spectra by using empirical line method (ELM) \ref{ELM}. Two more reference target surface with widely different brightness were required for this method; a white reference of $99.9\%$ reflectance which was placed with the sample and dark current acquired by closing camera's shutter. Using known reflectance values and acquired image radiance, the reflectance was calculated for each wavelength by following linear equation:

\begin{equation}
    R = \frac{R_r - B}{W - B}
\end{equation}
where $R$ is reflectance of data cube, $R_r$ is the radiance captured of given sample, $B$ and $W$ are the data captured for dark and white reference, respectively. Due to the variations in size of translational platform and the sample holder, Region of Interest (ROI) needed to be segregate from the acquired hyper-Cube. To automate the process of segregating region of interest from acquired hyper-cube, a false color image was created using bands from $450~nm$, $590~nm$ and $698~nm$ wavelengths. Furthermore, several contextual (pixel connectivity) and  non-contextual (thresholding) image segmentation techniques were applied to extract ROI. 

\subsection{Spectral Prepossessing}

Acquired spectral data is highly sensitive to physical properties of samples (temperature, surface , etc) and systematic noise (Ambient light, scattering, etc.). These noises can induce errors in acquired data and effect the reliability of build model. These errors can be avoided by standardizing the pure samples, which is often time consuming, may create artifact, expensive and sometimes physically impossible \cite{chemometrics}. However, there are mathematical techniques available in literature which can remove the effect of errors from spectral data. However, there has not been any method proven to be a standard for removing or avoiding such errors, rather several hit and try methods have been used to investigate which method suits best according to the nature of hyper-Cube. Therefore, in this work, pre-treatment techniques such as, savitzky golay filtering, standard normal variant (SNV), and multiplicative scatter correction (MSC) were applied to data separately before building model. Savitzky golay filtering was used for smoothing spectral data with eleven points and third order polynomial fitting. MSC and SNV are usually used to remove non-uniform scattering and effect of particle size \cite{multiplicative}. MSC works on mean spectrum of data while SNV works only on data. Hence for adulterated samples SNV was used, so that spectrum of adulterant and chili should not tangle while for pure samples, MSC was utilized to standardize the spectrum.

\begin{figure*}[hbt]
  \centering
  \includegraphics[scale=0.7]{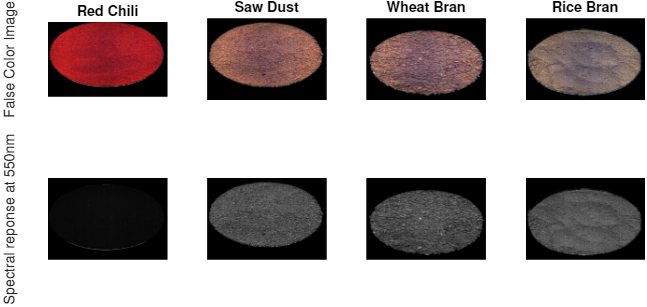}
  \caption{Spectral response of Red Chili and adulterants at $500~nm$}
\end{figure*}

 \begin{figure*}[h!]
\begin{subfigure}{.32\textwidth}
  \centering
  \includegraphics[scale=0.33]{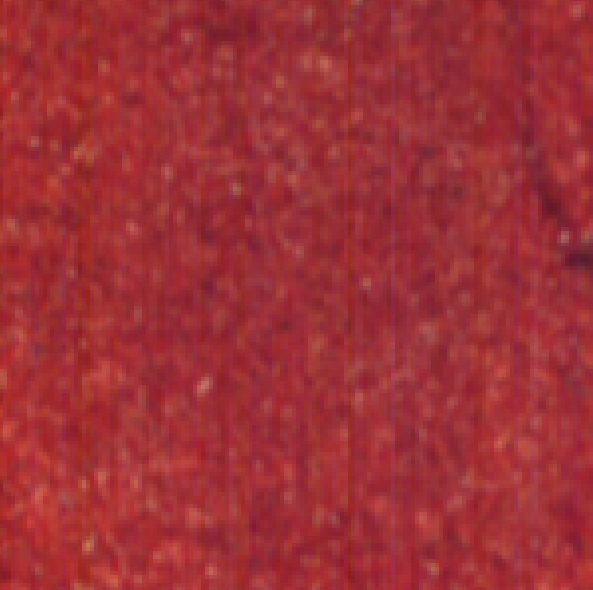}  
  \caption{False Color Image of adulterated Chili.}
  \label{Color}
\end{subfigure}
\begin{subfigure}{.38\textwidth}
  \centering
  \includegraphics[scale=0.33]{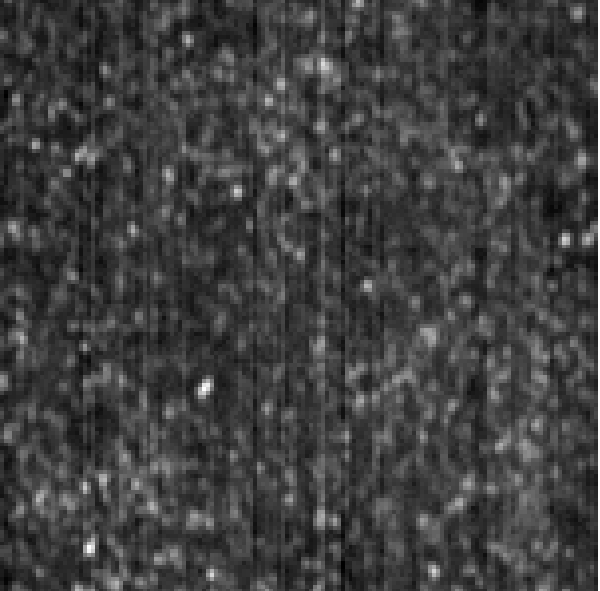}  
  \caption{Adulterated chili spectral response at $500~nm$.}
  \label{adultered}
\end{subfigure}
\begin{subfigure}{.3\textwidth}
  \centering
  \includegraphics[scale=0.33]{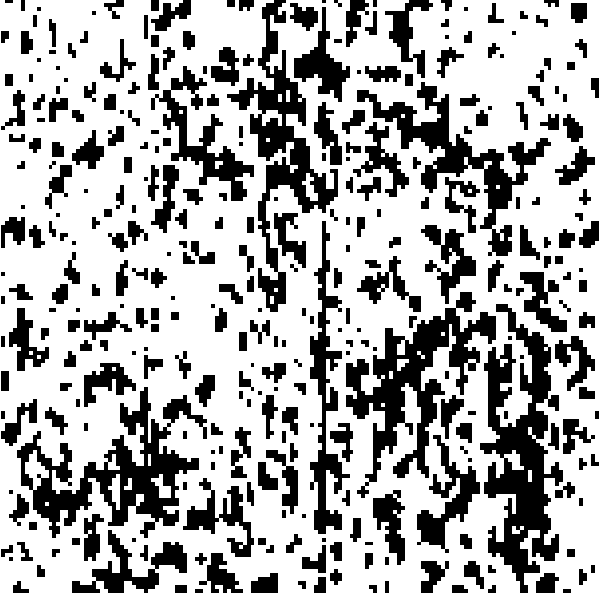}  
  \caption{Annotated Data}
  \label{labels}
\end{subfigure}
\caption{HSI data annotation process using K-Means clustering.}
\label{annotaion}
\end{figure*}

\subsection{Data Annotation}

To develop classification model, in order to differentiate red chili from adulterants and adulterants from each other, acquired spectrum of pure adulterants should be labeled. For labeling the data, this study exploit CIE standard (colorimetric) observer model (shown in Fig \ref{halogen}) which represent average human charomatic response. According to model, red color appears to be Dark (negative) from $450-550~nm$. As most of the digital camera used color filter array (CFL) and demosaicing algorithms to recover image \cite{kaur2015study}, this phenomena is not apparent in blue channel of image capture using digital camera. However, in HSI system hundred adjacent wavelength band can be acquired separately. Exploiting this feature of HSI system, one can observe and process the spectral response of the substance under consideration at specific wavelength. As red chili absorb blue wavelength due to its color, one can differentiate in red chili and adulterants with the acquired spectral response in blue wavelength. The most noticeable difference in red chili and adulterants reflectance is observe able at $500~nm$, hence image pixel at this wavelength can be grouped into two clusters.i.e. red chili \& adulterants.

 K-means clustering is one of the most popular clustering algorithm \cite{kmean}. It works on calculating the smallest distance of each data point from centroids and assigning it to nearest centroids. The best cluster center is selected by assigning data points to randomly chosen centroid and choosing cluster center again based on current data assignment \cite{Kmeans}. However, D Arthur et al. proposed that instead of choosing initial centroids randomly, the furthest point should be considered as initial centroid \cite{kmeans++}. For data annotation of this experiment, spectral image at $500~nm$ \ref{adultered} is fed to K-means algorithm with two cluster centers. The algorithm group image cells in two clusters based on their intensity values and assigned labels to each group member. Fig. \ref{labels} displays the labels assigned to each pixel where black color represent adulterants in mixture while white color is for red chili.
 
\subsection{Dimensionality Reduction}

Hyper-Cube is composed of hundreds of spectral bands covering a range of electromagnetic spectrum with very high spectral resolution narrow bands that not only improves the measurement capabilities of spectral cube but at the same time it brings challenges like storing, processing and classifying that data. With the increase in spectral bands/dimensions, the precision of classification decreases \cite{rojas2015curse}. It is due to the problem of finding and learning the structure of data embedded in high dimensional space as due to the rise in number of features the Hyper-cube has, the more data points are needed to fill the space. This phenomenon is known as the curse of dimentionality and to avoid this different approaches have been used in order to reduce the dimensions of hyper-Cube \cite{ma2013hughes}.

\begin{figure*}[!hbt]
\begin{subfigure}{.5\textwidth}
  \centering
  \includegraphics[scale=0.7]{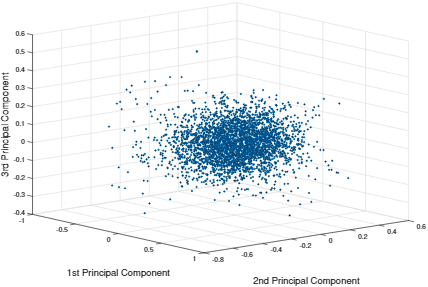}  
  \caption{Red Chili PCA: $85\%$ of total information }
  \label{redChiliPCA}
\end{subfigure}
\begin{subfigure}{.5\textwidth}
  \centering
  \includegraphics[scale=0.7]{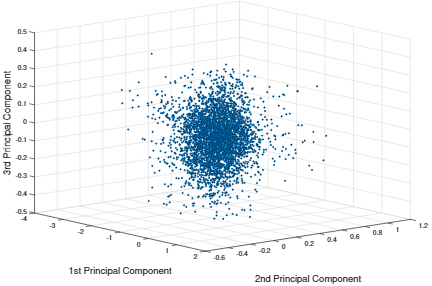}  
  \caption{Saw Dust PCA: $86\%$ of total information }
  \label{sawDustPCA}
\end{subfigure} \\ \linebreak

\begin{subfigure}{.5\textwidth}
  \centering
  \includegraphics[scale=0.7]{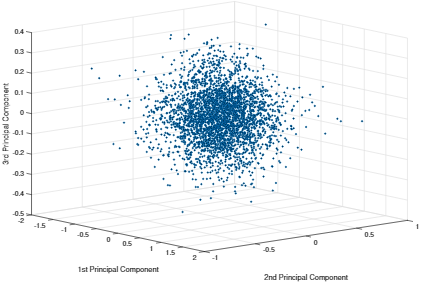}
  \caption{Wheat Bran PCA: $87\%$ of total information }
  \label{wheatBranPCA}
\end{subfigure}
\begin{subfigure}{.5\textwidth}
  \centering
  \includegraphics[scale=0.7]{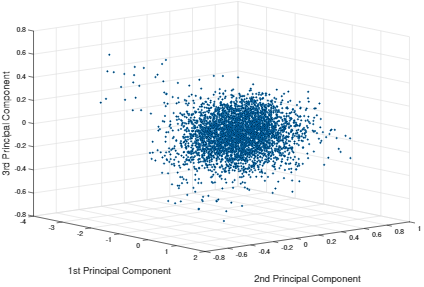}  
  \caption{Rice Bran PCA: $87\%$ of total information}
  \label{riceBranPCA}
\end{subfigure}
\caption{PCA score plot of the four materials: Red Chili, Saw Dust, Wheat Bran and Rice Bran.}
\label{PCA}
\end{figure*} 

Dimensionality reduction deals with complexities of large data set like hyper-Cubes and reduce its dimensionality by keeping the important features intact \cite{articleDimensionReduction}. Supervised methods like Linear Discriminate Analysis (LDA) \cite{balakrishnama1998linear}, Local Fisher Discriminate Analysis (LFDA) \cite{li2011locality} and unsupervised methods like Principal Component Analysis (PCA) \cite{gupta2006wavelet} and Maximum Noise Fraction transform (MNF) \cite{yokoya2010maximum} are mostly used in reducing hyperspectral dimensions by projecting the original data in a lower dimensional space. To reduce the dimension of hyper-cube data used in this study, PCA has been applied before further implementation.

In this study, Support Vector Machines (SVM) algorithm is used to identify the type of  adulterants in chili and as the number of classes, that needs to be classified, increases the numbers of parameters increases which in return affects the accuracy of the classification \cite{moughal2013hyperspectral}. Therefore, PCA has been used to avoid the curse of dimensionality and to ensure the quality of classifying adulterants. PCA ignores the spatial information of the data and reveals the internal structure in a way that most explains the variance in data. Orthogonal transformation projects the property of $224$ hyper bands, where the first $2$ projection of red chili, saw-dust, wheat bran, and rice bran contains $85\%$ \ref{redChiliPCA}, $86\%$ \ref{sawDustPCA}, $87\%$ \ref{wheatBranPCA} and $87\%$ \ref{riceBranPCA} of total variance respectively. Therefore, PC1 and PC2 has been selected in this study for classification purpose.

\subsection{SVM: One Class}

\begin{figure*}[!hbt]
\begin{subfigure}{.5\textwidth}
  \centering
  \includegraphics[scale=0.3]{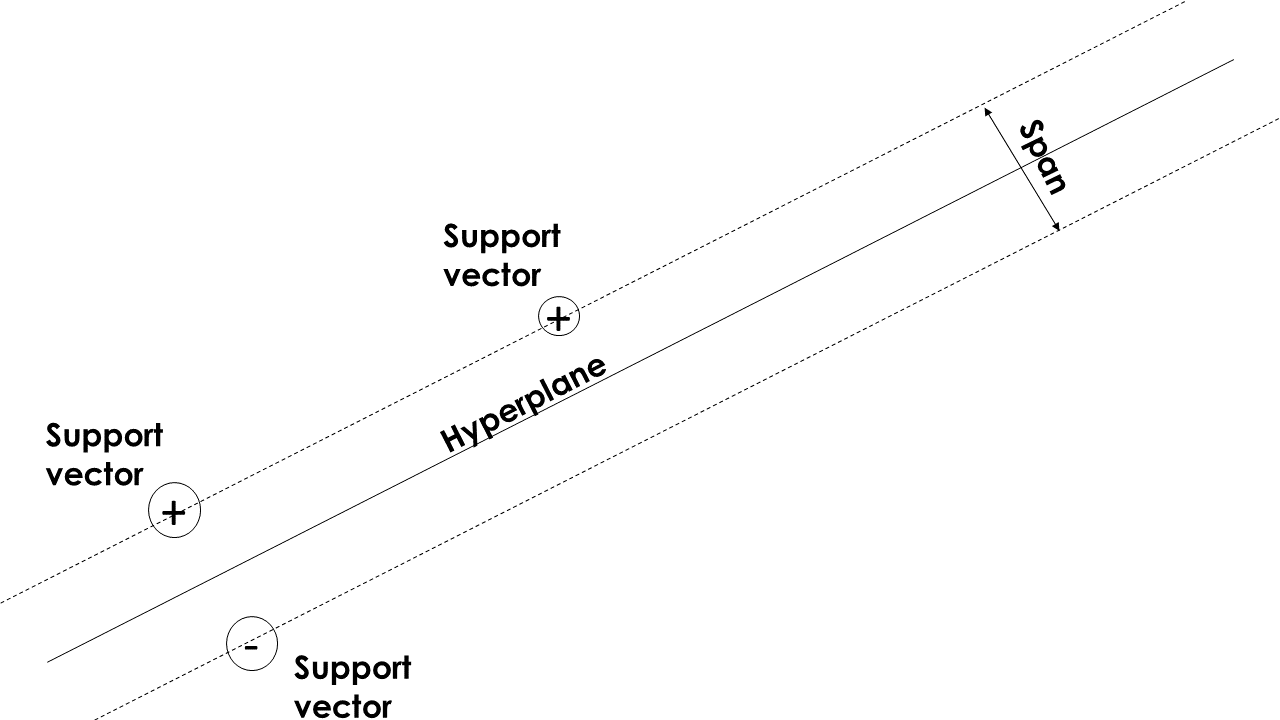}  
  \caption{Illustration of Support Vector Machine.}
  \label{SVM}
\end{subfigure}
\begin{subfigure}{.5\textwidth}
  \centering
  \includegraphics[scale=0.36]{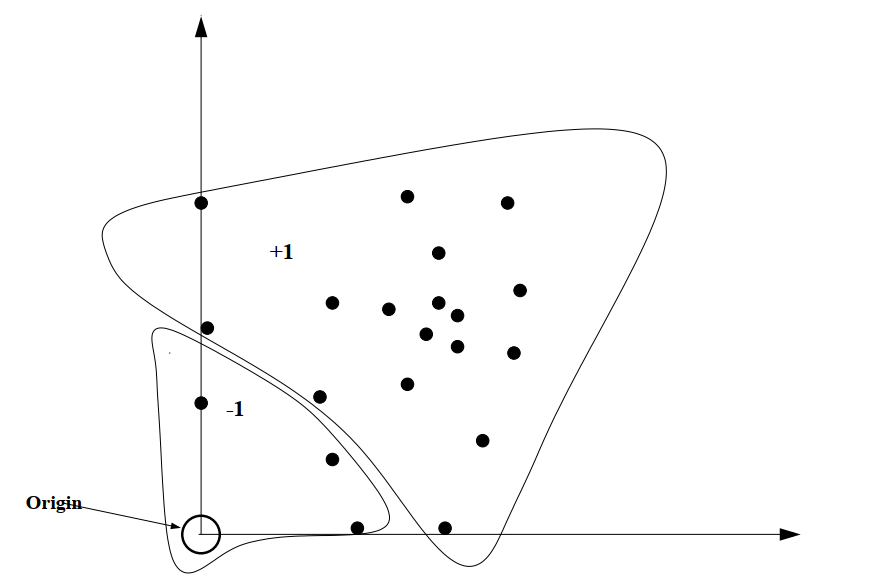}
  \caption{B. Schölkopf method for one class SVM \cite{2001one}}
  \label{OneClass}
\end{subfigure}
\caption{Support Vector Machine Concept.}
\label{fig:fig}
\end{figure*}

SVM is a supervised machine learning algorithm which map input data into feature space and draw a linear decision boundary \cite{support}. SVM in its original form was developed as binary classifier and could only assign two labels; $1$ and $-1$ to a given dataset. It classifies the data by taking into account only those training samples that lies on the boundary of the class distribution, known as support vectors, and identifying the optimal hyper-plane between two classes. Illustration of these definitions are given in Fig. \ref{SVM}.

SVM algorithm was initially designed for linear separable data, however, in real-life scenarios, the data is not always linearly separable. Boser, et al. proposed a method to create non-linear classifier by using kernel tricks \cite{boser}. They replaced each dot product with kernel function which transformed data to non-linear or high dimensional space. The classifier remained as hyper-plane in transformed space but its non-linear in input space. The choice of appropriate kernel and its parameters (width ($\gamma$), step size, regularization parameter ($C$), etc.) is dependent on the application-domain and type of training data. The larger value of $C$ and small $\gamma$ may lead to over fitting of model, other way around it may cause under fitting. Similarly, the higher value of $\gamma$ enlarge the area of support vector and also increase the elasticity of decision boundary while the smaller value of $\gamma$ increases maximum margin and decreases the flexibility of decision boundary. There is no single criteria to decide these parameters and the only approach is hit and trial \cite{determination}.

SVM in its true nature is a binary algorithm i.e. positive and negative examples are required to train an algorithm. B. Schölkopf et al., proposed a modification in SVM algorithm for one class classification problem. They use origin as the only member of second class and draw a hyper-plane to segregate class of interest from origin with maximal margin \ref{OneClass}. Therefore, in this study, one class linear (without feature transformation) as well as non-linear SVM with different kernels (polynomial, Gaussian and Gaussian radial based function (rbf))  are considered. For non-linear SVM kernel, kernel width ($\gamma$) is estimated using grid search which evaluate the model for all possible values hyper-parameters in specified range and proposed best suitable value. The detailed illustration of these settings is explained in results and discussion section.

\section{Results and Discussion}
\label{sec4}
\begin{figure}
\centering
  \includegraphics[scale=0.25]{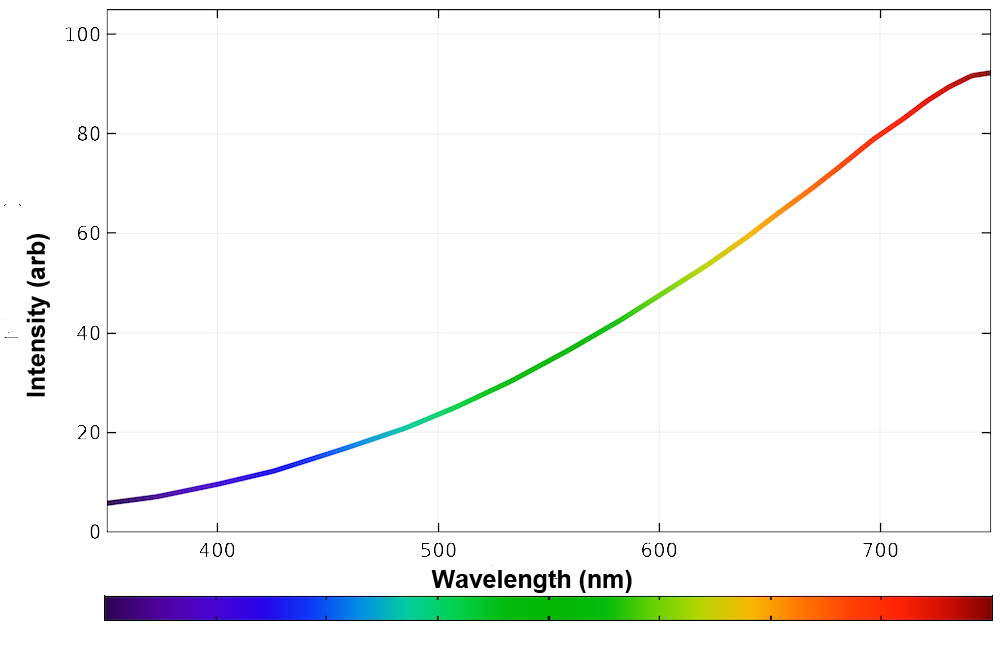}
  \caption{Halogen illumination Spectrum}
  \label{halogen}
\end{figure}

In this experiment, Halogen lamps are used for the illumination of samples. Although, halogen lamp shown in Fig. \ref{halogen} cover the whole range of HSI system ($400-1000~nm$) spectrum use in this study in contrast to LEDs and fluorescence tubes which lack continuous spectrum, but it has very low intensity of blue light \ref{halogen}. This limitation cause lower signal to noise ratio (SNR) in first $15$ bands, which can effect the performance of developed model. Therefore, initial $15$ bands are discarded and spectral range of $435-1000~nm$ is considered for experiments. As all pixels in a pure sample exhibit same characteristics spectra with negligible variations, therefore, each pixel is considered as a separate sample for model development. PCA is applied to spectral data of all pure samples to describe the characteristics features of spectra and number of important PC's are selected by broken stick method\cite{BS} which consider eigenvalue only if its value is greater than broken stick distribution.   

\subsection{Detection of Red Chili}

\begin{figure}[h!]
\centering
  \includegraphics[scale=0.42]{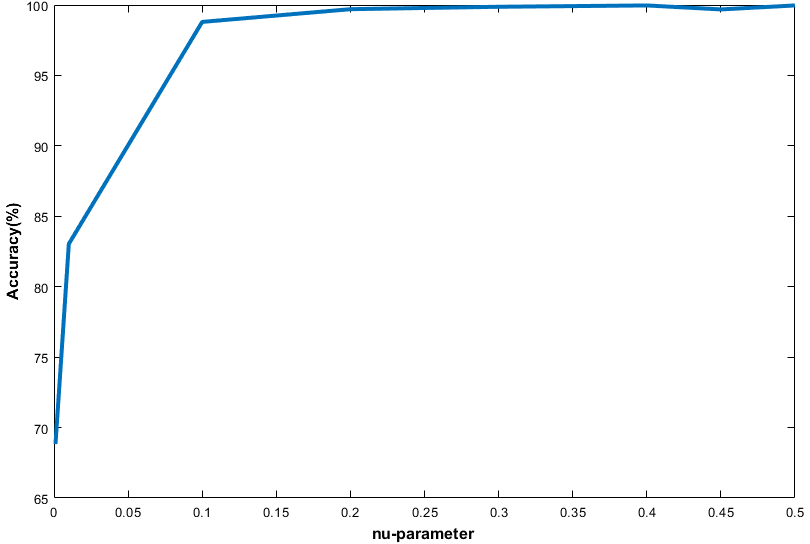}
  \caption{($\nu$)-parameter Vs. Accuracy}
  \label{nu}
\end{figure}

During the red chili detection, there is only a single class i.e. pure red chili, which do not need to be labeled. One class SVM algorithm with linear as well as non-linear kernels are trained on data. An important parameter $\nu$, which control the upper and lower limit of training error is need to be estimated as the higher value of $\nu$ parameter will not incorporate all training data. On other hand, very small value of $\nu$-parameter will also consider outliers in training data and decrease the testing accuracy. The value of $\nu$-parameter is estimated by hit and trial method while training has been done on pure red chili and for testing purpose pure adulterant and adulterated samples are used. Fig. \ref{nu} depicts that accuracy of classifier sharply decreased as $\nu$-parameter is set to be less than $0.1$. The accuracy is measured while testing with pure adulterants and minimum accuracy among all test samples is considered. Hence, for this experiment the value of $\nu$-parameter is considered to be $0.1$. 

A linear SVM model is trained on red chili spectral data with $\nu = 0.1$ and classifier predict different samples of red chili with an accuracy of $99~\%$. But when pure adulterants samples are fed to classifier for prediction, the classifier was unable to distinguish between red chili and adulterants. The maximum accuracy achieved using linear SVM is $14.8674~\%$ with $nu$-parameter value $0.89$. Therefore, non-linear one class SVM with $rbf$ kernel is considered. The optimal value of $\gamma$ is found to be $0.1$ with $\nu$-parameter $0.1$ using grid search method. Non-linear $rbf$ kernel is able to differentiate among chili and pure adulterants (rice bran, wheat bran and saw-dust) with an accuracy of $100~\%$. Similarly, polynomial kernel with $degree = 3$ is trained on data, where the value of $nu$-parameter is set to be $0.1$. The best accuracy value i.e $1.89~\%$ is achieved with $\gamma = 10$. This result depict that only one class SVM with $rbf$ kernel is able to differentiate red chili from adulterant with sufficient accuracy.

\begin{figure}
  \includegraphics[scale=0.85]{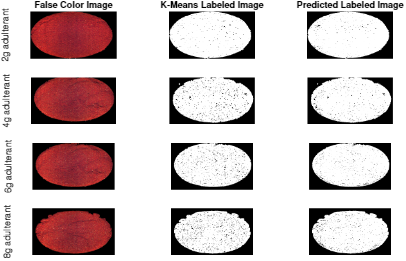}
  \caption{Red Chili adulterated with Rice Bran}
  \label{fig:RiceBran}
\end{figure}

\begin{figure}
  \includegraphics[scale=0.85]{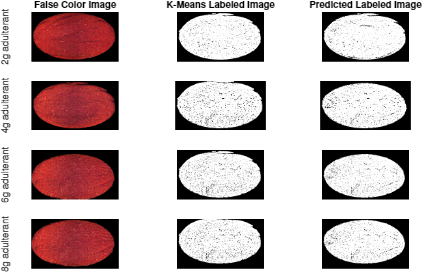}
  \caption{Red Chili adulterated with Saw Dust}
  \label{fig:RiceBran}
\end{figure}

\begin{figure}
  \includegraphics[scale=0.85]{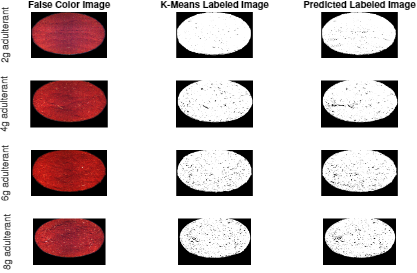}
  \caption{Red Chili adulterated with Wheat Bran}
  \label{fig:RiceBran}
\end{figure}

However, in case of adulterated samples, the efficiency is sharply reduced to $85~\%$. The efficiency of classifier decreased with the increased in adulteration due to the limitation in data annotation process as the penetration depth of blue light is $0.5-2.5~mm$ while IR wavelength can penetrate to a depth of $8-10~mm$ \cite{depth}. This indicates that grains below surface particles cannot be labeled with confidence, however, IR radiations can detect their properties. Similarly, due to smaller grain size, there exist several pixels which contains particles of both i.e. red chili \& adulterants. Although mixed pixel spectra is considered as anomaly, hence adulterants, in one class classification but data annotation process assigned labels based on the proportion of red chili and adulterants.

Another limitation faced during this experiment is the difference in densities of adulterants and red chili. Each gram of red chili and different adulterant have different bulk volume which results in different number of particles in acquisition process. As the HSI system do not incorporate density information explicitly and scan only the information of particles on the upper layer of sample, hence this troubles algorithm in predicting the accurate proportion of chili and adulterants. It is worth noting that all above mentioned limitations are related to data annotation process while the classifier predict pure adulterants samples with $100~\%$ accuracy.

\section{Conclusion}
\label{sec5}

In this research, one of the main challenge in spice industry i.e., adulteration in red chili, has been addressed. The HSI data acquired by VNIR HSI system has been pre-treated using savitzky golay filtering and standard normal variant (SNV). Data has been labeled by applying k-means clustering at $500~nm$ spectral response due to color feature of red chili. To reduce the dimensions of hyper-cube and to increase the classification accuracy, PCA was applied to spectral data before one class SVM classification. Overall, $99~\%$ accuracy has been achieved in case of pure red chili and a decreasing trend has been observed with the increase of adulterant in sample due to limitation in data annotation process and different wavelengths penetration depth. In order to detect adulterant types and estimate the adulteration proportion, spectral unmixing is a viable solution which will be exploited in further studies.

\bibliographystyle{ieeetr}
\bibliography{sample}
\end{document}